\documentclass[preprint,5p,11pt]{elsarticle}

\usepackage{amsmath, amssymb, amsfonts, bm}
\usepackage{graphicx}

\usepackage[dvipsnames]{xcolor}
\usepackage{kotex,soul}
\setstcolor{red}

\journal{Chaos, Solitons and Fractals}
\begin{document}

\begin{frontmatter}
\title{Reinforcement Learning Optimizes Power Dispatch in Decentralized Power Grid}

\author[1]{Yongsun Lee}
\author[1,2]{Hoyun Choi}
\author[3]{Laurent Pagnier}
\author[2]{Cook Hyun Kim}
\author[2]{Jongshin Lee}
\author[1]{Bukyoung Jhun}
\author[4]{Heetae Kim}
\author[2,5,6]{Jürgen Kurths}
\author[2]{B. Kahng\corref{cor1}}
\ead{bkahng@kentech.ac.kr}
\cortext[cor1]{corresponding author}

\affiliation[1]{
    organization={CTP and Department of Physics and Astronomy, Seoul National University},
    city={Seoul},
    postcode={08826},
    country={Korea}
}

\affiliation[2]{
    organization={CCSS, KI for Grid Modernization, Korea Institute of Energy Technology},
    city={Naju},
    postcode={58217},
    state={Jeonnam},
    country={Korea}
}

\affiliation[3]{
    organization={Department of Mathematics, The University of Arizona},
    city={Tucson},
    postcode={85721},
    state={Arizona},
    country={USA}
}

\affiliation[4]{
    organization={KI for Grid Modernization, Korea Institute of Energy Technology},
    city={Naju},
    postcode={58217},
    state={Jeonnam},
    country={Korea}
}

\affiliation[5]{
    organization={Potsdam Institute for Climate Impact Research},
    city={Telegraphenberg},
    postcode={D-14415},
    state={Potsdam},
    country={Germany}
}

\affiliation[6]{
    organization={Institute of Physics, Humboldt University Berlin},
    city={Berlin},
    postcode={D-12489},
    country={Germany}
}

\begin{abstract}
Effective frequency control in power grids has become increasingly important with the increasing demand for renewable energy sources. Here, we propose a novel strategy for resolving this challenge using graph convolutional proximal policy optimization (GC-PPO). The GC-PPO method can optimally determine how much power individual buses dispatch to reduce frequency fluctuations across a power grid. We demonstrate its efficacy in controlling disturbances by applying the GC-PPO to the power grid of the UK. The performance of GC-PPO is outstanding compared to the classical methods. This result highlights the promising role of GC-PPO in enhancing the stability and reliability of power systems by switching lines or decentralizing grid topology.
\end{abstract}

\begin{highlights}
\item The advent of renewable sources has introduced instability into modern power grids.
\item Maintaining the frequency of grid is imperative, even in the power plant accidents.
\item It is challenging to decide how much unbalanced power each bus should compensate.
\item We address the issue through deep reinforcement learning with graph neural network.
\end{highlights}

\begin{keyword}
    Reinforcement learning \sep Graph neural network \sep Decentralized power grid \sep Power dispatch
\end{keyword}

\end{frontmatter}

\section{Introduction}
As global warming accelerates, our energy systems must transition rapidly and substantially to those that utilize renewable energy sources, such as biomass, hydroelectric power, wind, and solar energy. Electric power grids, a fundamental component of this transition, are rapidly adopting renewable energy sources. However, as the proportion of renewable energy in the grid increases, grid systems will become more susceptible to instability~\cite{smith2022,Bottcher2022,Florian2013,Motter2013,schmietendorf2017,ulbig2014}.

Challenges abound in managing such hybrid power systems. Solar and wind energy generation depends heavily on daily and seasonal weather patterns. Furthermore, the electric frequency can fluctuate within seconds, owing to variations in sunlight and wind intensity caused by clouds and storms. Consequently, regulating electric currents from renewable sources is becoming increasingly vital for stabilizing these frequencies~\cite{Edenhofer2011,Milan2013,anvari2016,zhang2019,tyloo2020}. Additionally, renewable energy power plants are often far from consumer locations, necessitating long-distance transmission lines and posing a cascading failure risk~\cite{Menck2014,pesch2014}.

Therefore, optimizing decentralized hybrid power grids has emerged as a critical consideration. Power grids transmit two types of alternating current (AC): inertial AC, derived from fossil fuel combustion or nuclear fission, and inertia-free AC, generated by renewable sources and connected via power electronic inertia-free inverters~\cite{milano2018,anvari2020,schafer2022,tielens2016}.
External disturbances such as transmission line disconnections or regional overloads can disrupt the grid, with inertia-free AC unable to recover spontaneously, potentially leading to widespread frequency desynchronization and grid failure~\cite{simonsen2008,Menck2013,tyloo2019,hindes2019,schafer2018non}.

Considerable efforts have been made to reduce these dynamic disturbances and avoid large-scale power grid blackouts. Several methods have been proposed and implemented, such as controlling the time-dependent feedback (e.g., fast frequency responses~\cite{meng2019}), increasing the global inertia by connecting turbines without generators~\cite{alipoor2014,pagnier2019optimal} and switching off uncontrollable generators~\cite{Rebours2007}. A traditional method of recovering voltage or frequency is to add extra power saved in electric storage systems~\cite{heide2011,fleer2016} or request plants to produce extra power calculated using the optimal reactive power dispatch algorithm~\cite{schiffer2014,taher2019}. These methods have been effectively applied for a long time because the power grid is centralized, and the number of plants is small. However, in power grids composed of many small solar and wind plants, such strategies may not be optimal owing to their slow response and limited adaptability to rapid changes in power generation~\cite{Schafer2018,bottcher2020}. Thus, finding an optimal method to recover instantaneously or dispatch power to stabilize the grid system is a significant challenge.

In this context, reinforcement learning (RL) has emerged as a promising approach for devising optimal dynamic strategies. RL is often used to determine optimal dynamic pathways in various fields, ranging from the game of Go to autonomous driving~\cite{Silver2016,Sutton2018,Kwon2020,Mirhoseini2021,Degrave2022}, and provides efficient algorithms for diverse phenomena. Accordingly, we propose a novel power dispatch strategy using the RL approach, specifically the Graph Convolutional Proximal Policy Optimization (GC-PPO) algorithm. Using this method, we obtain information on the amount of extra power produced by each generator to minimize frequency fluctuations across the grid, offering quick adaptability to diverse grid configurations. The extra power supply of each generator is heterogeneous and depends on the topology of the power grid.

\section{Results}

\subsection{Swing equation of oscillators in the power grid}
We consider a power grid comprising $N_g$ generators and $N_c$ consumers. Therefore, the power grid comprises $N=N_g+N_c$ buses. Their phases and frequencies are denoted as ($\theta_i$, $\dot \theta_i$), with index $i=1,\cdots N$. The buses are connected to other buses via transmission lines and are treated as nodes in the graph representation. An oscillator at bus $i$ rotates according to the swing equation (also called the second-order Kuramoto model)~\cite{alexander1986,qiu2020}
\begin{equation}    \label{eq:swing}
m_i \ddot \theta_i + \gamma_i \dot \theta_i = P_i  + \sum_{j=1}^{N} K_{ij} \sin(\theta_j -\theta_i ),
\end{equation}
where $m_i$ is the angular momentum (or called inertia), $\ddot \theta_i$ is the angular acceleration, $P_i$ is the amount of power generation ($P_i >0$) or consumption ($P_i < 0$), $\gamma_i$ is the damping coefficient of oscillator $i$, and $K_{ij}$ is the coupling constant (or called line susceptance) between connected buses $i$ and $j$, and is given as $K_{ij}=|V_i||V_j|/x_{ij}$, where $V_i$ and $V_j$ are the voltages of buses $i$ and $j$, and $x_{ij}$ is the reactance of the transmission line. If buses $i$ and $j$ are not connected, then $K_{ij}=0$.

\begin{figure*}
\centering
\includegraphics[width=0.9\textwidth]{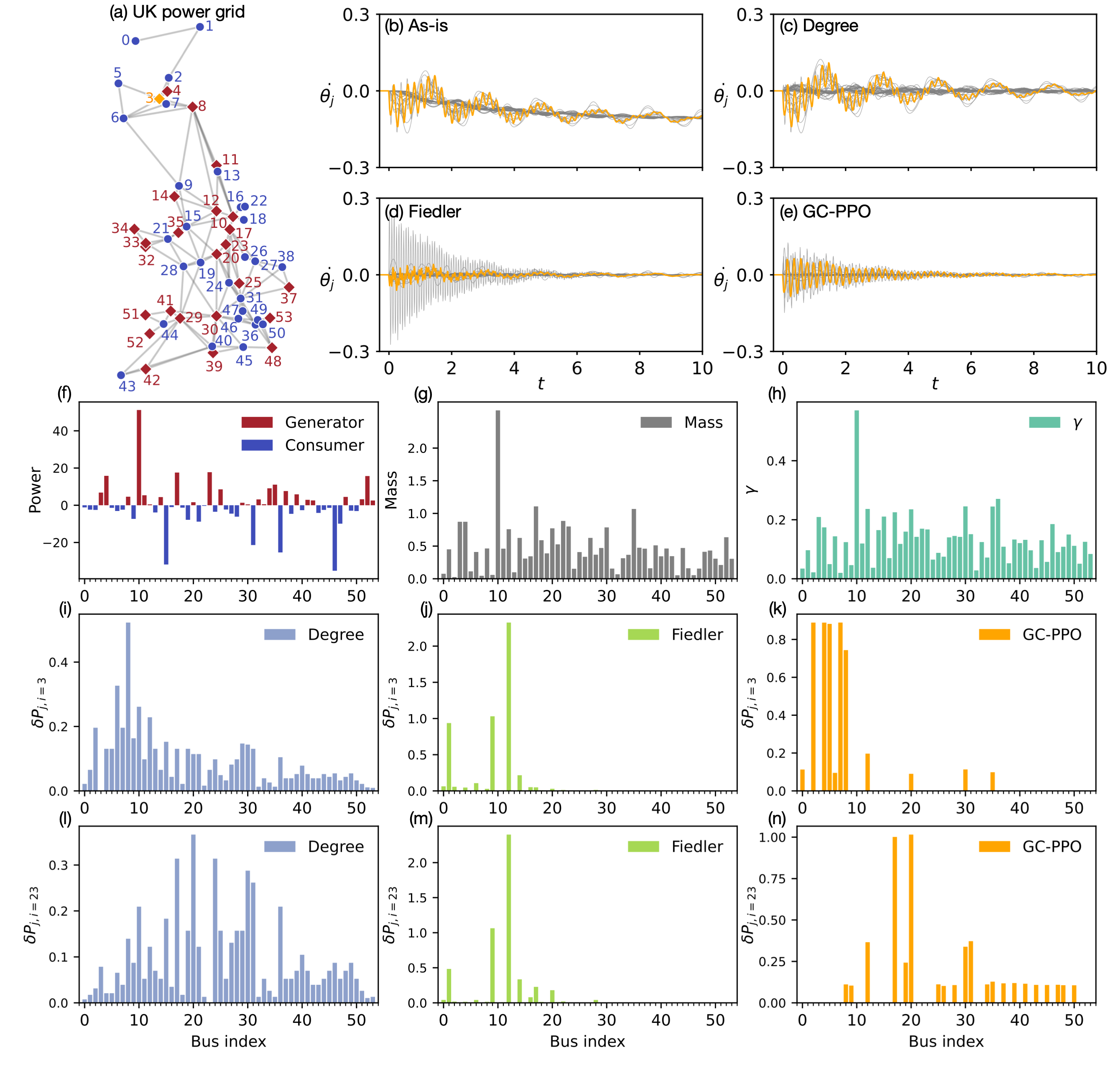}
\caption{
(a) Power grid of the high-voltage transmission lines in the UK coarse-grained by the Kron reduction method. Buses are composed of generators (Red diamond) with $P_i >0$ and consumers (blue circle) with $P_i<0$.
(b) The frequency relaxation pattern of each bus after bus 3 (yellow diamond) is perturbed, and no power dispatch is applied. The frequencies for the other buses are drawn in gray.
(c)$-$(e) Fluctuation relaxation pattern when the amount of power dispatch is determined by three different protocols: (c) Degree, (d) Fiedler, and (e) GC-PPO.
Here are plots of the (f) power, (g) mass (inertia), and (h) damping coefficient versus bus indices of the UK grid.
(i)$-$(k) Amount of power dispatch generated from bus $j$ when bus $i=3$ is perturbed, i.e., $\delta P_{j, i=3}$ for the three different protocols.
(l)$-$(n) Similar to plots (i)$-$(k), but when bus 23, which is located at the center of the grid, is perturbed.
}   \label{fig:fig1}
\end{figure*}

\subsection{Network models}
We test our method on different networks to validate it. Here, we use both synthetic SHK networks~\cite{Schultz2014} that have the advantage of being easily generated and a realistic model of the UK grid, which captures more faithfully the details of real-world power systems. The SHK model is designed to reflect the features of real-world grids from various aspects. The topology is governed by a trade-off between a tree-like structure to minimize costs and ensuring redundant backup lines for emergencies.

Physical variables in the SHK grid are taken as $m_i = m = 1$, $\gamma_i = \gamma = 0.5$, and power is chosen as $P_i=P=1$ (generators) and $P_i=P=-1$ (consumers). The proportions of generators and consumers are even, whereas their positions on the graph are randomly assigned. $K_{ij}=K=4$.

The real-world power grid of the UK comprises 235 buses, where generators control their power generation, and consumers regulate their consumption, for example, with fast frequency response. The power grid is reduced to 54 buses comprising $N_g=25$ renormalized generators and $N_c=29$ renormalized consumers using the Kron reduction method~\cite{dorfler2013} for computation. Since these reduced buses consist of multiple generators and consumers, their power $\{P_i\}$ can be controlled.
The topology of the reduced UK grid is illustrated in Fig.~\ref{fig:fig1}(a), where the red diamond buses represent generators ($P>0$) and the blue circles represent consumers ($P<0$). The power, mass, and damping coefficient of each bus are considered as their physical values (Figs.~\ref{fig:fig1} (f)$-$(h)). The UK grid comprises buses with different inertia, including renewable sources, and transmission lines with different coupling constants. As a result, it has more complex dynamics than the synthetic power grid.

\subsection{Bus-based power dispatch}   \label{subsec:bus_based_dispatch}
To determine the steady synchronous state of a given power grid, we first randomly select $\theta_i$ in the range $(-\pi, \pi]$ and $\dot{\theta}_i=0$ for each $i=1, \dots, N$. Then, the nodes' $\{\theta_i$, $\dot \theta_i\}$ values are updated following the swing equation~\eqref{eq:swing} until they reach a steady state. The power-dispatch process begins with the initial configuration of the steady state $\{ \theta_i, \dot\theta_i \}$. Next, we consider two situations where a stable power grid is perturbed: (i) when a generator malfunctions or a consumer overuses power. For failure in generator $i$, the reduced power generation $\delta P_i$ is set as, for example, $0.3 P_i$. In addition to the change in power, the inertia of generator $m_i$ is changed to $m_0 + 0.7(m_i-m_0)$. $m_0$ is set to $0.02$ for the SHK grid and $0.025974$ for the UK grid, which is the smallest inertia in the system. (ii) In the case of an overload, only the power consumption of the consumer $i$ increases by $\delta P_i = 0.3 \vert P_i \vert$, but the inertia $m_i$ remains unchanged. Finally, to maintain the power balance $\sum_k P_k = 0$, each remaining bus $P_j$ ($j\ne i$) must generate additional power denoted by $\delta P_{ji}$, which satisfies $\sum_{j \in \mathcal{B} \backslash \{i\}} \delta P_{ji} = \delta P_i$. Here, $\mathcal{B}$ is defined as the set of all buses. $\delta P_{ji}$ can be determined according to various protocols. The details are described in METHOD section~\ref{sec:methods}.

\begin{figure}
\centering
\includegraphics[width=0.99\linewidth]{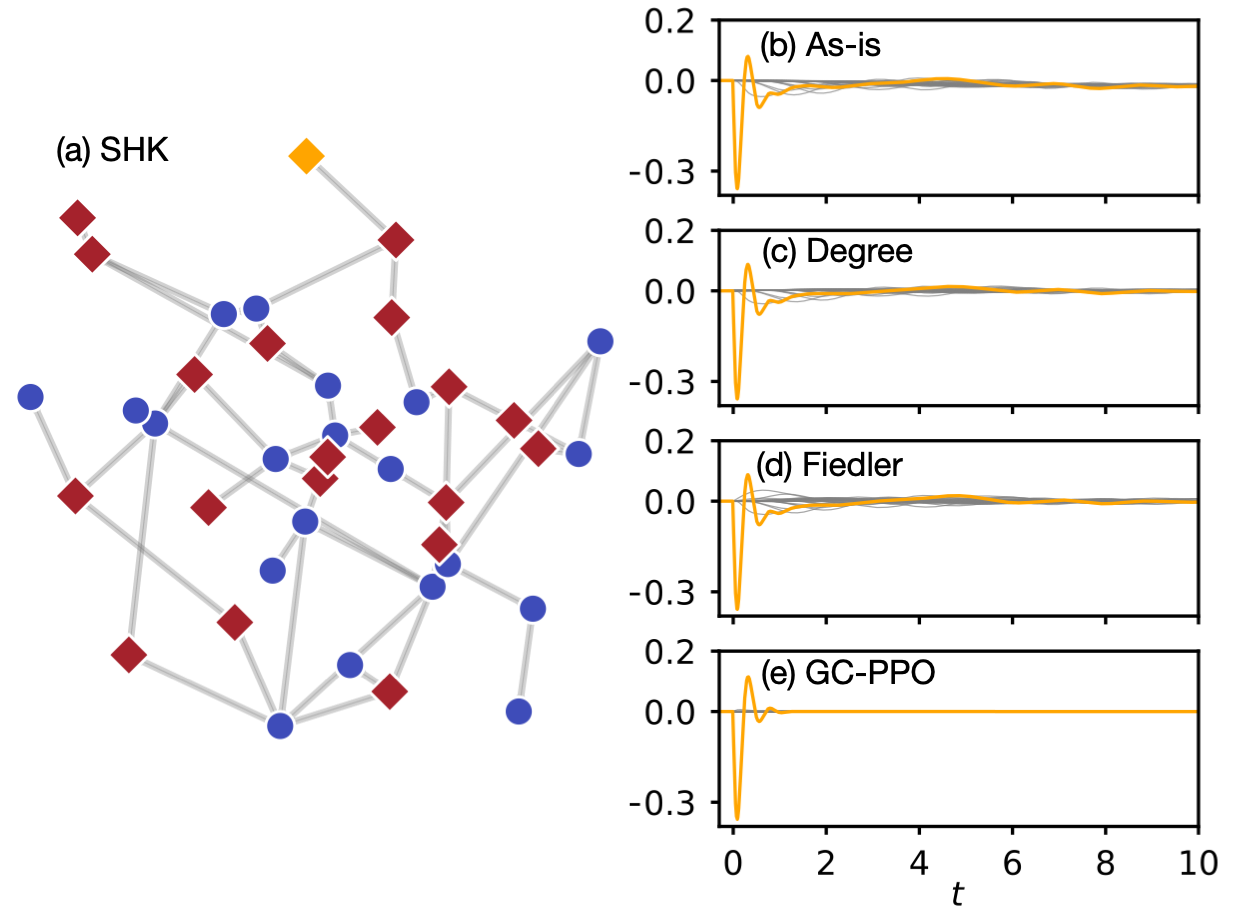}
\caption{
(a) Topology of the synthetic SHK grid, composed of generators (Red diamond) with $P_i=1$ and consumers (blue circle) with $P_i=-1$.
(b) Frequency evolution of the perturbed bus (Yellow line, denoted by Yellow diamond in (a)) and the other buses (gray lines) without power dispatch.
(c)$-$(e) Fluctuation relaxation pattern when power dispatch is performed following three protocols: (c) Degree, (d) Fiedler, and (e) GC-PPO.
}
\label{fig:fig2}
\end{figure}

Fig.~\ref{fig:fig1}(b) shows the evolution of the frequency of each bus when a perturbation is applied at $t=0$ to generator 3, which is marked by a yellow diamond in Fig.~\ref{fig:fig1}(a), of the UK grid without power dispatch (As-is). Bus 3 exhibits the most severe fluctuation when no power dispatch is performed. The yellow line indicates the frequency of the perturbed bus 3, whereas the gray lines are the frequencies of the other buses. Figs.~\ref{fig:fig1}(c)$-$(e) show the frequency patterns of bus 3 and the other buses after power dispatch is applied following the Degree, Fiedler, and GC-PPO protocols, respectively. The value of $\delta P_{j, i=3}$ is given according to each protocol, as shown in Figs.~\ref{fig:fig1}(i)$-$(k). Because all buses except the perturbed one participate in the power dispatch, only the disturbed value $\delta P_{i, i}$ is zero.
Although all protocols reduced fluctuations, the pattern of fluctuation relaxation suggests that the GC-PPO protocol achieves the best improvement among the protocols we tested.

Figs.~\ref{fig:fig1}(l)$-$(n) show plots similar to those shown in Figs.~\ref{fig:fig1}(i)$-$(k), when generator 23, located almost at the center of the grid, is perturbed. Each protocol responds differently, depending on perturbation $\delta P_i$. As shown in both Figs.~\ref{fig:fig1}(k) and (n), GC-PPO dispatches more power to generators in the neighborhood of the perturbed bus $i$, although it does not impose any constraints on the distance. In contrast, the Degree protocol redistributes power relatively evenly among the other generators. The Fiedler protocol focuses on a few buses (1, 9, 12) regardless of the perturbed bus.

Figs.~\ref{fig:fig2} are corresponding plots to Figs.~\ref{fig:fig1}(a)$-$(e), but the network is adopted from the synthetic SHK power grid. Since the SHK grid is more homogeneous than the UK grid in topology, masses, and links aspects, the fluctuations of the buses are not so large compared to those of the UK grid. Nevertheless, when GC-PPO protocol is employed, the system quickly recovers the stable steady state.

\begin{table*}
\centering
\begin{tabular}{c|cc|cc}
                & \multicolumn{4}{c}{Fluctuation measure $\Xi$}                                                                                     \\
\hline
Power grid   & \multicolumn{2}{c|}{SHK}  & \multicolumn{2}{c}{UK}                                                            \\
\hline
Perturbation & Generator                 & Consumer                  & Generator                 & Consumer                  \\
\hline
As-is        & $5.25 \times 10^{-3}$     & $4.92 \times 10^{-3}$     & $1.71 \times 10^{-1}$     & $1.48 \times 10^{-1}$     \\
Uniform      & $4.29 \times 10^{-3}$     & $4.12 \times 10^{-3}$     & $3.89 \times 10^{-2}$     & $2.92 \times 10^{-2}$     \\
Degree       & $4.32 \times 10^{-3}$     & $4.19 \times 10^{-3}$     & $3.85 \times 10^{-2}$     & $2.92 \times 10^{-2}$     \\
BC           & $4.43 \times 10^{-3}$     & $4.31 \times 10^{-3}$     & $3.73 \times 10^{-2}$     & $3.27 \times 10^{-2}$     \\
Clustering   & $6.64 \times 10^{-3}$     & $5.74 \times 10^{-3}$     & $4.30 \times 10^{-2}$     & $2.87 \times 10^{-2}$     \\
Fiedler      & $6.26 \times 10^{-3}$     & $6.19 \times 10^{-3}$     & $4.58 \times 10^{-2}$     & $5.98 \times 10^{-2}$     \\
GC-PPO       & \bf 2.54 $\times 10^{-3}$ & \bf 2.34 $\times 10^{-3}$ & \bf 1.13 $\times 10^{-2}$ & \bf 1.07 $\times 10^{-2}$ \\
\hline
\end{tabular}
\caption{
Total fluctuation measures $\Xi$ over all buses for different protocols for the synthetic network and real-world UK power grid. Note that a smaller value of $\Xi$ indicates a better protocol. The GC-PPO protocol is more efficient for the more heterogeneous UK power grid.
}
\label{tab:protocols}
\end{table*}

\subsection{Fluctuation measure}   \label{subsec:measure}
If power dispatch is successfully implemented, all buses will be synchronized with the previous frequency, and the grid will be stabilized. Otherwise, buses would not be in a synchronized state, and a global blackout could occur. Therefore, the stability of the power grid must be measured accurately after dispatch~\cite{witthaut2022,Mitra2017}.

Here, we introduce $\Xi_i$, a fluctuation measure of the grid after the power is dispatched owing to a perturbation on bus $i$.
\begin{equation}    \label{eq:measure}
    \resizebox{0.91\hsize}{!}{
        $\Xi_i \equiv \frac{1}{T} \int_0^T dt \left[ \frac{1}{\sum_k m_k} \sum_k m_k \dot \theta_k^2 (t) -\left( \frac{1}{\sum_k m_k} \sum_k m_k \dot \theta_k (t) \right)^2 \right]$
    }
\end{equation}
The fluctuation measure $\Xi_i$ is the weighted variance of the frequencies $\dot \theta_k (t)$ over all buses, including the generators and consumers. The weight is taken as inertia $m_k$~\cite{paganini2017}. The frequencies are monitored for $T$ seconds after the initial perturbation at $t=0$.

As the generator or consumer that causes the unexpected perturbation is not recognized, we introduce the average $\Xi_i$ over all possible perturbations $\Xi \equiv \sum_{i \in \mathcal{G}} \Xi_i / N_g$ for the generators and the $\sum_{i \in \mathcal{C}} \Xi_i / N_c$ for the consumers.
Naturally, the smaller $\Xi$ is, the more stable the power grid and the better the protocol performance.

\begin{figure}
\centering
\includegraphics[width=0.99\linewidth]{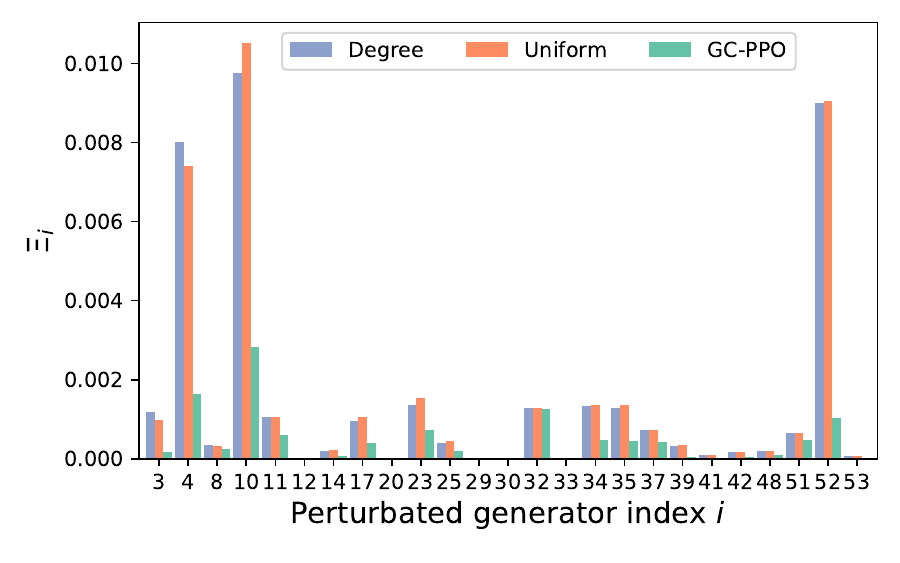}
\caption{
Fluctuation measure $\Xi_i$ versus generator index $i$ in the UK grid for three protocols: Degree, Uniform, and GC-PPO.
Degree and Uniform protocols perform poorly on three generators (4, 10, and 52), while GC-PPO effectively moderates the fluctuations.
}
\label{fig:nodewise_perf}
\end{figure}

Fig.~\ref{fig:nodewise_perf} shows the fluctuation measure $\Xi_i$ versus the perturbed generator index $i$ for the three protocols: Degree, Uniform, and GC-PPO. For example, for $i=3$, $\Xi_i$ of the GC-PPO is obtained based on the fluctuating data $\dot{\theta}(t)$ shown in Fig.~\ref{fig:fig1} (e). While the Degree and Uniform protocols result in a similar pattern based on the peaks of $i=4$, 10, and 52, the GC-PPO protocol produces significantly smaller fluctuations. Generator 4 is in a sparsely connected region, and generator 52 is located on a leaf, making them topologically vulnerable nodes. Bus 10 is a generator with large $P_i, m_i, \gamma_i$ as shown in Figs.~\ref{fig:fig1} (f)$-$(h) and therefore, the perturbation caused by $0.3 P_i$ significantly affects the stability of the entire system.
Overall, the new GC-PPO protocol is more effective than the other protocols in reducing the instability of the UK power grid.

The $\Xi$ values are listed in Table.~\ref{tab:protocols}, for six protocols to compare their capabilities with the raw values obtained without any protocol. In the case of the SHK grid, the fluctuation is small even when power dispatch is not performed (As-is). The physical quantities such as power, mass, damping coefficient, and coupling constants are rather homogeneous. Nevertheless, GC-PPO yields $\Xi$ values several times smaller than the other protocols. GC-PPO demonstrates a considerably more dramatic performance improvement for the UK grid with heterogeneous physical quantities than for the synthetic grid.
Whereas the other protocols exhibit $\Xi$ improvements of approximately 10 times over no power dispatch (As-is), GC-PPO reduces $\Xi$ by more than 100 times.

\subsection{Training of GC-PPO}

\begin{figure*}
\centering
\includegraphics[width=0.8\linewidth]{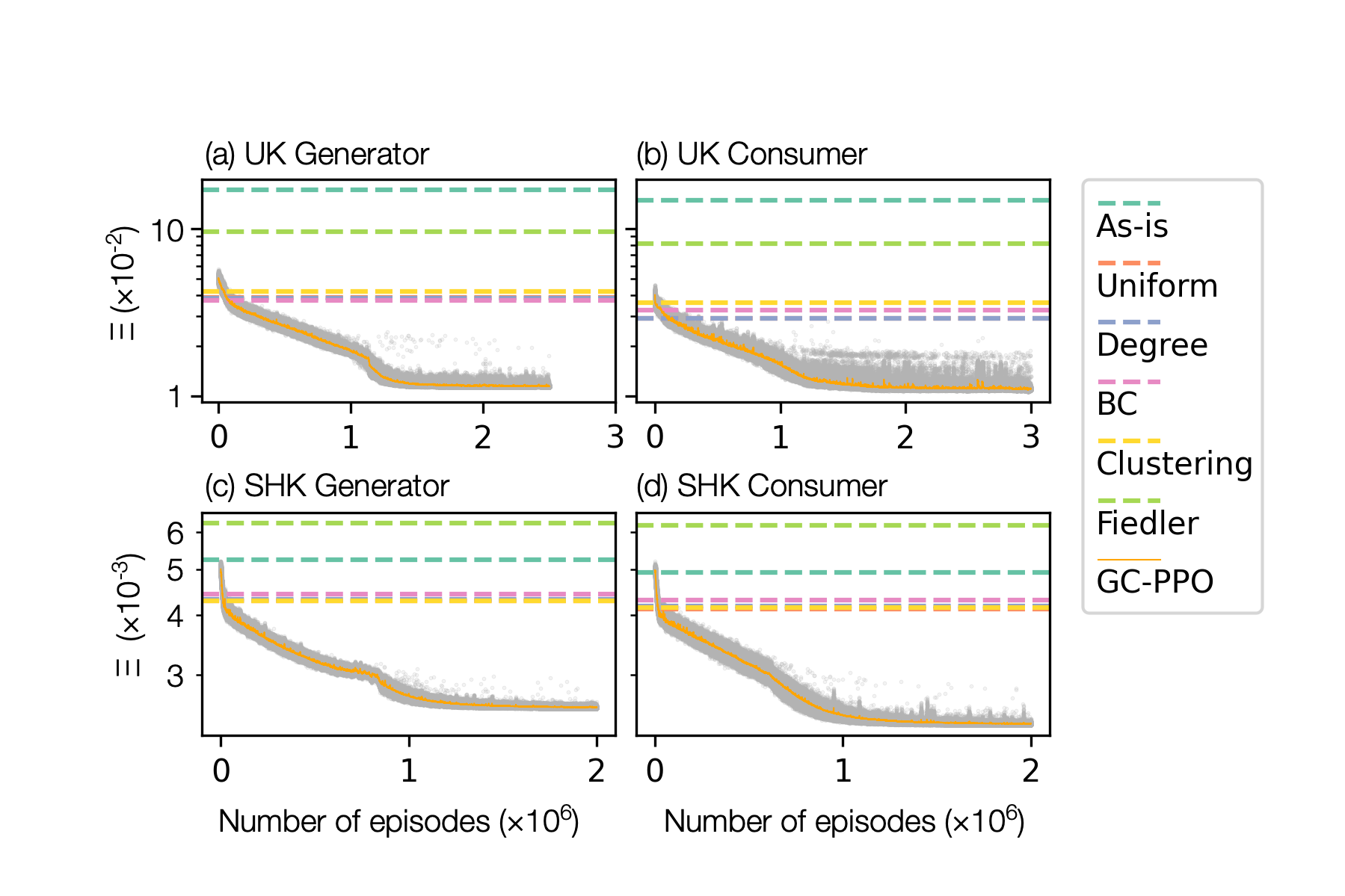}
\caption{
Performance of GC-PPO versus the number of training episodes in different environments for the different power grids and bus types. The solid yellow curves indicate the average performance of GC-PPO, whereas the gray dots represent the fluctuations due to the stochastic training process. The dashed lines are the performances of the other topology-based protocols for reference.
}    \label{fig:training}
\end{figure*}

In contrast to the other protocols, the new GC-PPO protocol requires training.
During training, the GC-PPO learns by experiencing various power dispatches and refining the method it uses to compute $\delta P_{j, i}$. While the detailed training method is described in Section~\ref{sec:methods}.B, observing the progress of the GC-PPO performance would be beneficial. Fig.~\ref{fig:training} shows a decrease in $\Xi$ as the training of GC-PPO progresses. The dotted lines represent the $\Xi$ values for other protocols in Table.~\ref{tab:protocols}. Because the other protocols do not require training, their $\Xi$s values remain constant regardless of the training episodes of the GC-PPO.
Across all power grids and perturbations, the GC-PPO clearly outperforms the other protocols, with a quick drop {of $\Xi$} in the early stage of training.
It then shows a gradual improvement in performance until it reaches $10^6$ training episodes.
Finally, $\Xi$ reaches a plateau where the improvement is negligible, indicating that the GC-PPO has been sufficiently trained.

\begin{figure}[tb]
\centering
\includegraphics[width=0.99\linewidth]{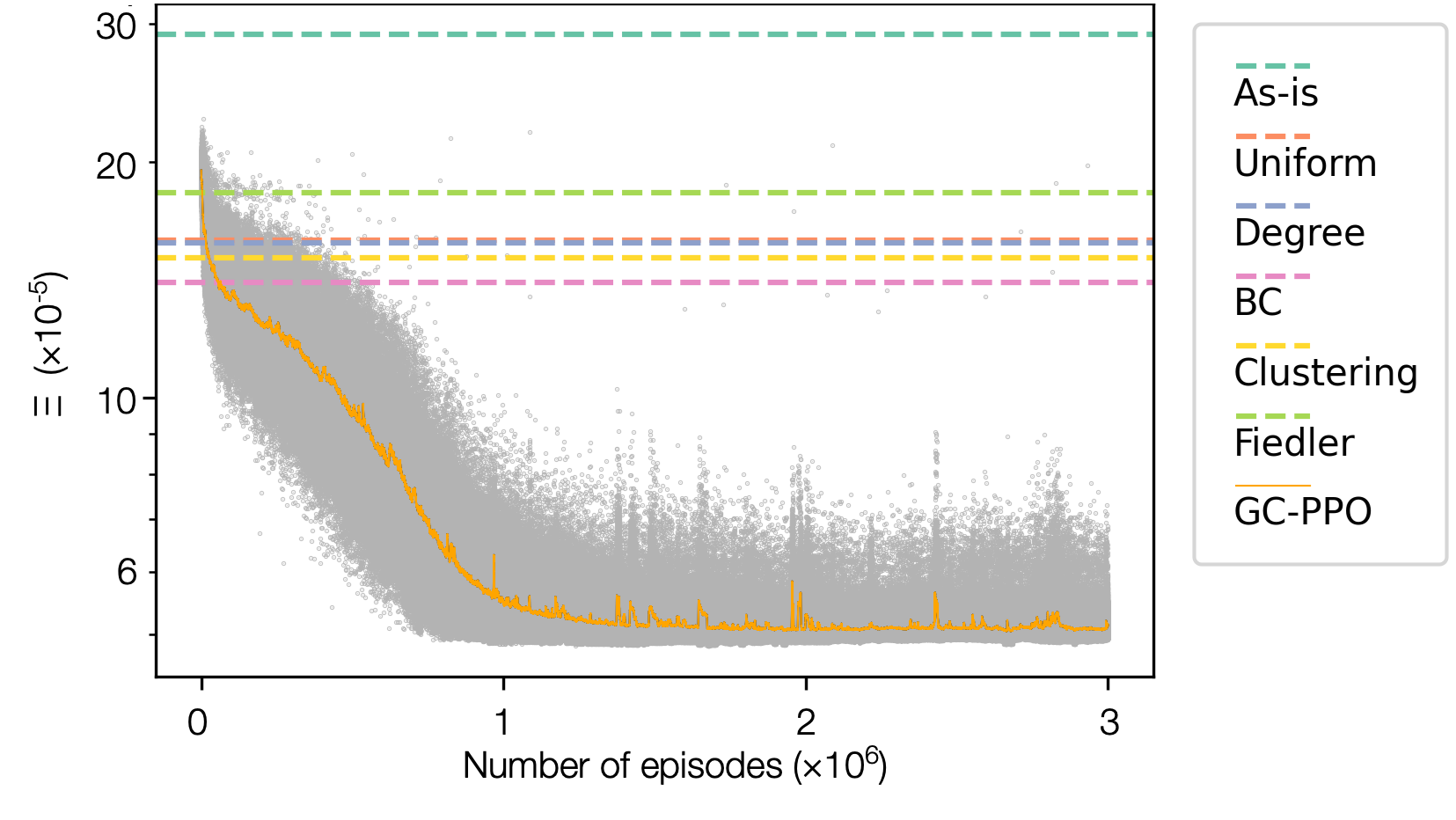}
\caption{
Training of the GC-PPO when multiple consumers simultaneously overuse power. Solid and dashed lines are equivalent to Fig.~\ref{fig:training}.
} \label{fig:training_mult}
\end{figure}

Power shortages may occur when heaters or coolers are used simultaneously in a town or region because of significant weather changes. We envision a situation where six consumers in the northern part of the UK grid with bus indices 0, 1, 2, 5, 6, and 7 overuse power simultaneously. Fig.~\ref{fig:training_mult} shows the training curve for GC-PPO performance under this perturbation. Similar to Fig.~\ref{fig:training}, this figure reveals that GC-PPO outperforms the baseline protocols after a few training episodes, and its performance gradually improves over approximately $10^6$ training cycles.

\section{Discussion}    \label{sec:discussion}
We developed a novel optimal power-dispatch method using the GC-PPO algorithm classified into reinforcement learning. This method determines how much power {\it each bus} should dispatch a power to compensate for the lost power to stabilize the system. This method is compared to the traditional method of supplying the lost power from a few power plants. This paradigm shift in the power dispatching method has been a timely need as the power grid has become decentralized due to the proliferation of small-scale solar and wind power plants. The innovative algorithm of GC-PPO introduced here outperforms the classical method.

The GC-PPO algorithm optimizes a variation measure $\Xi$, which is the weighted variance of the frequency over all buses. The weights are the inertia $m_k$ of each bus $k$. This weighted strategy is more effective than the unweighted strategy in obtaining additional power $\delta P_{j,i}$, particularly for a heterogeneous power grid, meaning generators with larger inertia can be asked to produce more power to recover the system quickly. On the other hand, for the SHK model, we assume uniform inertia for all buses, i.e., $m_i=m=1$. Then, no weights are needed to measure fluctuations.

Notably, GC-PPO is not limited to the specific situation we have covered; to implement the algorithm in other grid systems, we first pre-train GC-PPO on an ensemble of small power grids and then fine-tune it for a given large power grid~\cite{Nauck2022,Yang2021,nauck2023}. Perturbations are not necessarily limited to a single bus but also occur when multiple consumers overuse simultaneously.
Furthermore, GC-PPOs can also be proactive in grid stability if they are informed about possible faults~\cite{Jhun2023}.

Changing the grid's topology by switching lines is an alternative to maintain the stability of the power grid.
The next challenge will be to modify the GC-PPO algorithm to solve the switching problem.

\section{Methods}  \label{sec:methods}
Here, we describe the determination of the power dispatch $\delta P_{ji}$ owing to perturbation $\delta P_i$ for each protocol listed in Table.~\ref{tab:protocols}. Because the power balance relation $\sum_{j \in \mathcal{G} \backslash \{i\}} \delta P_{ji} = \delta P_i$ should be satisfied, we choose $\delta P_{ji}$ as proportional to $\delta P_i$.
\begin{equation}    \label{eq:regulation_topo1}
    \delta P_{ji} = \frac{q_{ji}}{\sum_{k \in \mathcal{G} \backslash \{i\}} q_{ki}} \delta P_i.
\end{equation}
Thus, each protocol determines $q_{ji}$ for generator $j$.

\subsection{Heuristics methods}   \label{subsec:topology_protocol}
We propose five protocols based on the structural characteristics of the power grid: uniform, degree, betweenness centrality (BC)~\cite{Freeman1977,Freeman1978}, clustering coefficients~\cite{Watts1998}, and spectral properties (Fiedler mode)~\cite{Pagnier2019}, which are important for assessing electric power grid vulnerability~\cite{degreeBC}. These approaches possess different properties. We use the shortest-path distance $d_{ji}$ between the perturbed bus $i$ and compensating bus $j$. Because the flux per unit length decays by $1/d$ with distance $d$ from a source in two dimensions, we choose $q_{ji}$ as inversely proportional to $d_{ji}$.

One of the simplest power dispatch protocols is the uniform protocol. All non-perturbed generators are treated equally, thus $q_{ji}=1/d_{ji}$. Note that this protocol uses only distance information from the perturbed bus.

The degree of bus $i$ ($k_i$) is defined as the number of buses connected to it via transmission lines. A bus with a large degree can exhibit increased stability when it is well-balanced with neighbors.
Therefore, we defined a degree protocol $q_{ji}=k_j/d_{ji}$.

The BC of bus $i$ ($b_i$) represents the number of times a bus $i$ is involved when each pair of buses transmits information along the shortest path between them. The higher the BC, the more frequently the random signal is received, which means that nodes with high $b_i$ are likely to be used for a detour when necessary and have more chances of being vulnerable.
The BC protocol is defined as $q_{ji}=b_j/d_{ji}$.

By contrast, the clustering coefficient of node $i$ ($c_i$) represents the amount of resilience against perturbations. A node with a higher clustering coefficient is less likely to be affected by frequency fluctuations in the electrical grid. Therefore, we define the clustering protocol as $q_{ji}=c_j/d_{ji}$.

Finally, Fiedler's mode is the eigenmode corresponding to the smallest nonzero eigenvalue in the Laplacian matrix and is defined as follows:
\begin{equation}
    L_{ij} =
    \begin{cases}
        -K_{ij} \cos(\theta_i^* - \theta_j^*)       & \text{if } i \neq j, \\
        \sum_j K_{ij} \cos(\theta_i^* - \theta_j^*) & \text{if } i = j.
    \end{cases}
\end{equation}
The Fiedler mode can be viewed as a set of sensitive buses, with amplitudes representing their sensitivities. Because they respond largely, even to a small external perturbation, it is expected that buses engaging in this mode tend to easily fluctuate around their stable points, which can cause the failure of another bus, eventually leading to a global cascading failure. Therefore, reinforcing the Fiedler mode $a_i$ amplitudes may be an effective method. In general, the mode's amplitude can be negative; therefore, we use it after taking the squared value as $q_{ji} = a_j^2 / d_{ji}$.

\subsection{GC-PPO protocol}    \label{subsec:RL_protocol}

\begin{figure}
\centering
\includegraphics[width=0.99\linewidth]{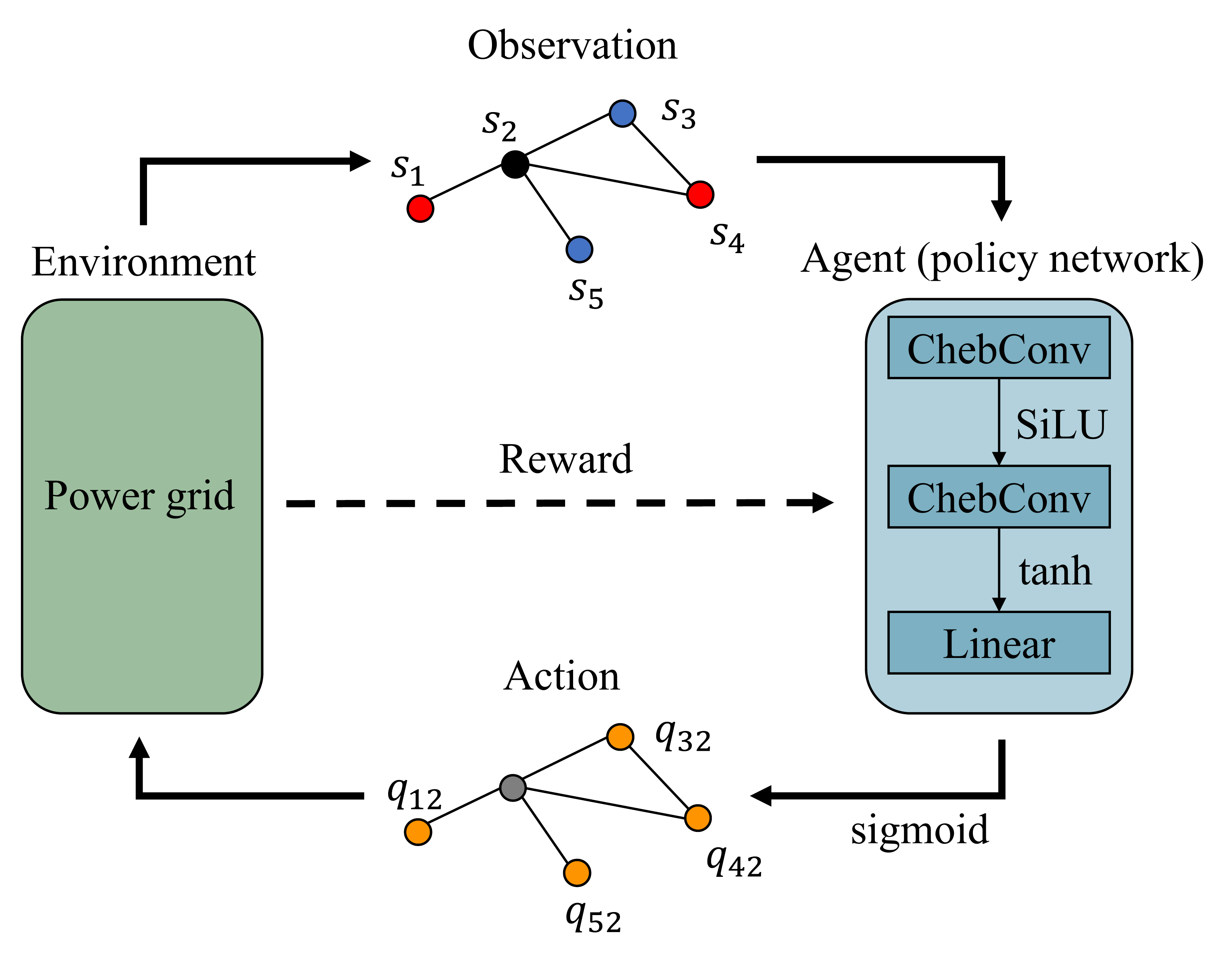}
\caption{
Feedback loop consisting of interactions between the environment (power grid) and agent (policy network). First, the agent observes the steady state and perturbation where generator 2 fails. Second, the agent outputs the action $q_{ji}$ using the observation. This neural network consists of two Chebyshev convolution layers with SiLU and tanh activations and a linear layer with sigmoid activation to constrain the output between 0 and 1. Lastly, the power grid is rebalanced, and the policy network is updated to maximize the reward.
} \label{fig:RL_step}
\end{figure}
Although classical topological protocols can capture the structural properties of a power grid, they cannot capture the temporal dynamics of a system. Designing a power-dispatch protocol that considers both complex topology and nonlinear dynamics driven by the swing equation~\eqref{eq:swing} is challenging. We propose a reinforcement learning (RL) approach called the Graph Convolutional Proximal Policy Optimization (GC-PPO) protocol to address this issue.

The RL scheme has two main components, the environment and the agent, composed of a power grid and a neural network called a policy network. The agent observes the environment and takes action based on its observations (Fig.~\ref{fig:RL_step}). The environment provides feedback (reward) on the outcome of the action to the agent to help make better decisions. The agent repeats its interactions with the environment and learns the optimal action for a given observation through trial and error.

Various RL algorithms have been proposed to train agents, and we employ one of the most robust algorithms, the PPO~\cite{Schulman2017}. This was proposed to avoid learning instability caused by fluctuations in the dynamical states by forcing the agent to take actions that are not significantly different from its previous actions during training. In addition, a policy network comprising the agent is implemented using graph convolution to understand the topological properties of the power grid. Notably, after appropriately modifying the structure of a neural network~\cite{Defferrard2016,Jiang2018}, we can effectively use a single agent for various networks.

GC-PPO training starts with the initial configuration $\{ \theta_i, \dot\theta_i \}$ in a synchronous state, as defined in Section.~\ref{subsec:bus_based_dispatch}. A perturbation on bus $i$ is given randomly, causing the entire system to lose its power balance. The current states $\{P_i, m_i, \gamma_i, \theta_i, \dot \theta_i \}$ of all the nodes, coupling constants $\{ K_{ij} \}$, and perturbation $\delta P_i$ are provided as inputs to the neural network.

For the policy network, we use Chebyshev convolution~\cite{Defferrard2016}, which considers both the Laplacian of the graph and its higher-order contributions. The first Chebyshev convolution layer with SiLU activation lifts the observations from the environment to a high-dimensional vector (Fig.~\ref{fig:RL_step}). The second Chebyshev convolution operation, activated by Tanh, computes a high-dimensional node-embedding vector. Next, all the node vectors share the linear and normalization layers and output a scalar value. Finally, the resulting scalar values are passed through a sigmoid function, returning the action $q_{ji} \in [0, 1]$ for the unperturbed bus $j$. Both Chebyshev layers have identical structures in terms of the number of convolution filters (four filters, considering the power of the Laplacian matrix up to $L^3$).

However, the output $q_{ji}$ from the policy network is not used directly for power dispatch $\delta P_{ji}$ as in Eq~\eqref{eq:regulation_topo1}. The GC-PPO protocol behaves differently when the agent is trained and is deployed after training. We consider a Bernoulli distribution with probability $q_{ji}$ for each unperturbed bus $j$ in the training phase. Buses are independently sampled from each distribution to determine whether to participate in the power dispatch. In other words, we set $q_{ji}$ of the nonparticipating bus to zero. This sampling process is crucial for training the agent to perform an optimal action~\cite{ishii2002,castronovo2013}. It enables exploration where the agent experiences diverse situations, even with the same $q_{ji}$, thereby building robustness into the policy.
Note that using the Bernoulli distribution is not the only method to force a probabilistic action.
One might sample $q_{ji}$ from a Dirichlet distribution parameterized by $\{\alpha_j\}_{j \in \mathcal{G} \backslash \{i\}}$, where $ \alpha_j \in [0, \infty]$.
However, the optimization of the agent suffers from the large parameter space of the Dirichlet distribution, making it practically impossible.
In contrast, the Bernoulli distribution with a reduced parameter space allows the neural network to be trained efficiently~\cite{Li2023}.

Meanwhile, deterministic action is preferred in the deployment phase to avoid unpredictable outcomes in critical infrastructures such as power grids. Therefore, we forgo the sampling process and set a threshold $q_\text{th}=1/10$, where $q_{ji}$ is taken as zero if it is less than $q_\text{th}$.

Thus, the output of the policy network is appropriately modified to $\delta P_{ji}$ following Eq~\eqref{eq:regulation_topo1}. After the power dispatch, the fluctuation $\Xi_i$ of the grid is measured for $T$ seconds. We set $T=2$ in the training phase to reduce the computational cost, while $T=10$ is used in the deployment phase to measure the performance of the GC-PPO precisely. As the agent is trained to earn more rewards, $-\Xi_i$ is passed on as a reward, completing one feedback loop. As discussed above, an initial failure can occur on any bus. Therefore, we repeat this feedback loop for every possible $i$, which we define as an episode.

The agent is trained following the standard PPO method as it progresses through one episode. For each feedback loop initiated by the perturbation of bus $i$, we compute the probability $p_j$ that the contributing generators will be selected. This is tractable because the generators are sampled in a Bernoulli distribution according to $q_{ji}$, which is the output of a policy network. Subsequently, the objective of loop $L_i$ is defined as follows:
\begin{equation}    \label{eq:loss}
    L_i= - \max \left[\frac{p_i}{p_i^\prime} \Xi_i, \textrm{clip} \left(\frac{p_i}{p_i^\prime}, 1-\epsilon, 1+\epsilon \right) \Xi_i \right],
\end{equation}
where $p_i^\prime$ is the probability calculated using the previous policy network.

The clip function limits the absolute value of the ratio of the probabilities calculated by the current and previous policy networks to $\epsilon = 0.1$, which is given as follows:
\begin{equation}    \label{eq:clip}
    \mathrm{clip}(x, x_\text{min}, x_\text{max}) =
    \begin{cases}
        x_\text{min} & \text{if}  \ x < x_\text{min},                    \\
        x            & \text{if}  \ x_\text{min} \le x \le x_\text{max}, \\
        x_\text{max} & \text{if} \  x_\text{max} < x.
    \end{cases}
\end{equation}
Finally, we average the objective $L_i$ over the entire episode and perform a gradient ascent over the averaged objective to update the policy network.

\section*{Acknowledgments}
This study was supported by the National Research Foundation of Korea No. RS-2023-00279802 (BK) and No. NRF-2022R1C1C1005856 (HK) and a KENTECH Research Grant No. KRG2021-01-007 (BK) and No. KRG2021-01-003 (HK).

The authors declare no conflicts of interest.

The code is available upon request.

\bibliographystyle{elsarticle-num}
\bibliography{main.bib}

\end{document}